\documentclass[preprint,aps,tightenlines,superscriptaddress,floatfix,
showpacs,nofootinbib]{revtex4}
\usepackage[final]{graphicx}
\usepackage{epsfig}

\usepackage{bm}

\newcommand{\eqref}[1]{({\ref{#1}})}

\begin{document}


\title{Parity nonconserving two-pion exchange in elastic proton-proton scattering}

\author{J.~A.~Niskanen}
\email{jouni.niskanen@helsinki.fi}
\author{T.~M.~Partanen}%
\email{tero.partanen@helsinki.fi}
\affiliation{
Department of Physical Sciences, P.~O.~Box~64, FIN-00014 University
of Helsinki, Finland}
\author{M.~J.~Iqbal}
\affiliation{ Department of Physics and Astronomy, University of
British Columbia, Vancouver, BC, Canada, V6T 1Z1}

\date{\today}

\begin{abstract}
Parity nonconserving two-pion exchange in elastic $\vec{p}p\,$
scattering is investigated in the presence of phenomenological
strong distortions in various models. Parity violation is included
in $NN\pi$ vertex considering $NN$ and $N\Delta(1232)$ intermediate
states in box and crossed box diagrams. Using the derived parity
nonconserving two-pion exchange potential we calculate the
longitudinal analyzing power $\bar{A}_{\rm L}$ in
elastic $pp$ scattering. The predicted effect is of the same order
as vector meson exchanges.
\end{abstract}


\maketitle

\section{\label{sec:level1}INTRODUCTION}

Although the basic weak interaction is relatively well known, there
exist large uncertainties of the weak couplings of mesons and
baryons at the hadron level relevant at low and intermediate
energies up to about 1 GeV, {\it i.e.} in nuclear physics. For
example the efforts to determine the parity nonconserving (PNC)
couplings \cite{ddh,fcdh} report uncertainties of the order of 100 \%
or more of the recommended "best" values. (For reviews see {\it
e.g.} Refs. \cite{ad,desrev}.) In this situation even the signs of
couplings may be suspect.

Considering the weakness of the couplings their experimental
determination is a big challenge and attempts often utilize nuclear
structure to enhance PNC effects. However, the analysis of the basic
couplings is then complicated by the environment. Among few-nucleon
systems also polarized photoreactions have been used with the
deuteron.

In principle, the most direct way without external disturbances
would be $NN$ scattering. Presently PNC $np$ scattering experiments
are unlikely and even $pp$ scattering experiments are scarce
\cite{kis,eve,los,ber}. Among these of particular interest is the
measurement of the parity violating spin observable $\bar A_{\rm L}$
in the TRIUMF experiment E497 at 221.3 MeV \cite{ber}. The energy
was chosen so that incidentally the strong interaction phases
conspire to cause the $J=0$ contribution to cross zero
\cite{simonius}. While in this amplitude both $\rho$ and $\omega$
mesons are equally important, the next $J=2$ parity mixing amplitude
is strongly dominated by $\rho$ exchange  enabling a drive towards
PNC $\rho pp$ coupling.

The above logic is based on the idea that the dominant PNC effect in
$pp$ scattering should be due to vector mesons, since long-ranged
single pion exchange is forbidden by Barton's theorem \cite{bar}.
This theorem forbids in general neutral $0^{\pm}$ mesons to couple
with nucleons in the PNC interaction because of the simultaneous
violation of $\mathcal{P}$ and $\mathcal{CP}$ symmetries. So in the
PNC $pp$ interaction at low energies only  $\rho$-, and
$\omega$-meson exchanges are expected to be significant and the
two-pion exchange (TPE) is assumed negligible \cite{ad}. With these
assumptions the TRIUMF experiment E497 \cite{ber} would, in
principle, result in the determination of the weak $\rho pp$
coupling constant
 $h_{\rho}^{pp}=h_{\rho}^{(0)}+h_{\rho}^{(1)}+h_{\rho}^{(2)}/\sqrt{6}$
from $pp\,$ scattering. The lower energy experiments \cite{kis,eve}
have already determined the independent combination of
  $h_{\rho}^{pp}$+$h_{\omega}^{pp}$
(where $h_{\omega}^{pp}=h_{\omega}^{(0)}+h_{\omega}^{(1)}$) and so
both the $h_{\rho}^{pp}$ and $h_{\omega}^{pp}$ are supposedly
determined separately.

In view of such dedicated and very time consuming experiments it is
important to carefully study the validity of the assumptions and
uncertainties in their interpretation. In fact, it was shown in Ref.
\cite{iq} that $\Delta$-isobar excitation by weak $\rho$ and strong
(mainly) pion exchange has a significant effect on $\bar A_{\rm L}$
at all energies and should also be considered. In spite of PNC
single pion exchange being excluded in $pp$ scattering, crossed
charged two-pion exchange is allowed. Charged pions are also
possible in two-pion exchanges involving the excitation of the
$\Delta^{++}(1232)$ resonance. The former was studied already in the
early works \cite{des,ris}, while calculations for the latter were
performed above the pion production threshold in Refs.
\cite{silbaretco,kloetco}. However, the $\Delta$ contribution
extends also below this to low energies (as shown with vector meson
exchanges in Ref. \cite{iq}). To our knowledge the contribution of
all these two pion effects to PNC observables has not been
investigated systematically on the same footing together. Closest
comes the recent Ref. \cite{kaiser} in deriving PNC the two-pion
potential in chiral perturbation theory. Our aim in the present
paper is to calculate this potential including realistic form
factors and extending the calculation to the observable $\bar A_{\rm
L}$. Further interest in TPE lies in the fact that it should be the
longest ranged PNC effect in $pp$ scattering, which might show up in
energy dependence.

The organization of the paper is as follows. First we evaluate
numerically PNC TPE as a potential in the momentum representation
assuming the baryons to be static. This potential can be well fitted
by Lorentz functions of the momentum transfer, which in turn can be
expressed in terms of Yukawa functions in the coordinate
representation. It may be noted that the result will be local
(excluding relativistic corrections) and a comparison with the local
parts of vector meson exchanges is then straightforward. Next we
discuss non-static effects in three different kinematics and dynamics
allowing kinetic energies for baryons. These are found to be
significant but not necessarily dominant.

\section{\label{sec:level2}THEORY}

\subsection{Basic interactions}

The Hamiltonians for the parity conserving (PC) $\pi NN$, $\pi
N\Delta$, and PNC $\pi NN$ interactions, which describe  the needed
nonrelativistic vertices are
\begin{eqnarray}
\label{h1} \mathcal{H}_{NN\pi}^{\rm PC}&=&
\frac{if_\pi}{m_\pi}N^\dagger\bm{\sigma}\cdot\bm{q}\bm{\tau}\cdot\bm{\pi}N,\\
\label{h2} \mathcal{H}_{N\Delta\pi}^{\rm PC}&
=&\frac{if_\pi^{\ast}}{m_{\pi}}\Delta^\dagger\bm{S}\cdot\bm{q}\bm{T}
\cdot\bm{\pi}N+\mbox{ h.c.},\\
\label{h3} \mathcal{H}_{NN\pi}^{\rm PNC}&=&\frac{h_\pi^{(1)}}
{\sqrt{2}}N^\dagger(\bm{\tau}\times\bm{\pi})_0N.
\end{eqnarray}
In this calculation only the $NN$ and $N\Delta$
intermediate states are considered and $\Delta\Delta$ intermediate
state ignored, because the PNC $N\Delta\pi$ vertex is concluded to
be very small in Ref. \cite{hen}. We take it to be zero.

\begin{figure}[tb]
\includegraphics[width=11.cm]{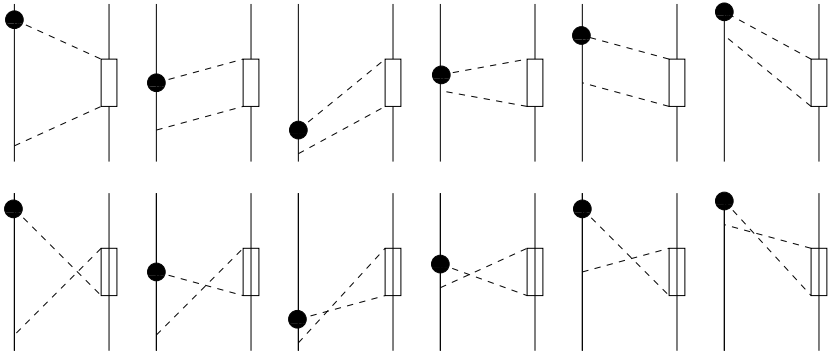}
\caption{\label{grf} Time orderings of parity nonconserving  (black
spot) two-pion exchange (dashed line) via nucleon (solid line) delta
(bar) intermediate state. Both the crossed and box graphs include
the $N\Delta$ intermediate state while the $NN$ intermediate state
can appear only in the crossed graphs. The direction of time is
upwards.}
\end{figure}

Unfortunately, in literature choices of the signs in the above
definitions vary, both overall and even in the $\pi NN$ {\it vs.}
$\pi \Delta N$ vertices. In purely strong interactions the signs
cancel, but in the presence of a weak vertex the sign matters. Our
choice follows the standard one (DDH \cite{ddh}) so that our PNC OPE
potential
\begin{equation}\label{weakpipot}
V_{NN\pi}^{\rm PNC}(\bm{r})=\frac{h_\pi^{(1)}f_\pi}{\sqrt{2}m_\pi}
(\bm{\tau}_1\times\bm{\tau}_2)_0(\bm{\sigma}_1+\bm{\sigma}_2)
\cdot\hat{\bm{r}}\frac{\partial}{\partial r}Y_\pi(\bm{r})
\end{equation}
would be as given in {\it e.g.} Ref. \cite{ad}. Here $Y_\pi(\bm{r})$
represents the Yukawa function $Y_\pi(\bm{r})=e^{-m_\pi r}/4\pi r$
(possibly modified by a form factor).

The change of the $\pi NN$ coupling to the $\pi \Delta N$ would be a
replacement of the nucleon spin-isospin operators $\bm{\sigma}$ and
$\bm{\tau}$ by the transition operators $\bm{S}$ and $\bm{T}$ with
the normalization
$S_i^\dagger{S}_j=(2\delta_{ij}-i\epsilon_{ijk}\sigma_k)/3$
\cite{bro}.

\begin{figure}[tb]
\includegraphics[width=10.cm]{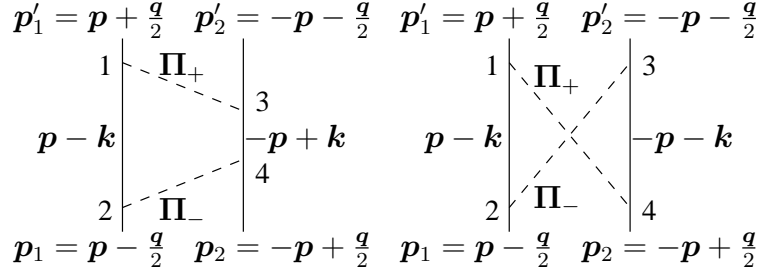}
\caption{\label{kin}The symmetry of the kinematics for  the box and
crossed graphs, where $\bm{q}$ is the momentum transfer, $\bm{k}$
the free loop momentum, and $\bm{p}$ the relative momentum between
the nucleons. The pions carry the momentum
$\bm{\Pi}_{\pm}=\bm{k}\pm\frac{\bm{q}}{2}$. Numbers 1-4 denote
respectively the number of the vertex.}
\end{figure}

Time orderings of the PNC TPE via $N\Delta$ intermediate state  can
be grouped under eight different groups, which all contain six time
orderings that have the same spin and isospin structure. Two of the
groups correspond to the graphs in Fig. \ref{grf}, where the upper
row is for the box (B) and the lower for the crossed (C) graphs, in
which the parity is broken in vertex 1 (the vertices are numbered in
Fig. \ref{kin}). The rest of the groups can also be formed by using
the graphs in Fig. \ref{grf}: Two of the groups include the same
graphs except the parity is broken in vertex 2. The remaining four
groups can be obtained in the same manner from Fig. \ref{grf} with
the exchange $N\leftrightarrow\Delta$ and breaking the parity
respectively in vertices 3 and 4. Time orderings of the PNC TPE via
$NN$ intermediate state correspond only to the crossed graphs in
Fig. \ref{grf}. This contribution has four different groups, ${\it i.e.}$
the parity is broken once in each vertex. Overall there are totally
72 time-ordered graphs to add up.

A symmetric and practical choice of the meson mediated momentum
transfers is $\bm{\Pi}_{\pm}=\bm{k}\pm\frac{\bm{q}}{2}$ shown in
Fig. \ref{kin}, where $\bm{k}$ is an integration parameter and
$\bm{q}$ the overall momentum transfer.

\subsection{\label{static}Static model}
In our basic model the initial and final nucleons and all  the
intermediate baryons are considered static. The energy (mass)
difference between the isobar and nucleon is denoted by
$\delta=M_\Delta-M$ and the energies of the exchanged pions are
$w_{\pm}=(m_{\pi}^2+\bm{\Pi}_{\pm}^2)^{\frac{1}{2}}$. The energy
denominators
of the different graph types given in Fig. \ref{grf} act as
propagators of the 'old-fashioned' time ordered second-order
perturbation theory
\begin{widetext}
\begin{eqnarray}\label{props}
\mathcal{D}_{N\Delta}^{\rm B}(\bm{k}, \bm{q})&=&
-\frac{1}{4w_+w_-}\Bigl\{\frac{1}{\delta}
\Bigl[\frac{1}{w_++\delta}+\frac{1}{w_+}\Bigr]
\Bigl[\frac{1}{w_-+\delta}+\frac{1}{w_-}\Bigr] \nonumber \\
&~&~~~~~~~~~~~~~~+\frac{1}{w_++w_-}\Bigl[\frac{1}
{w_+(w_-+\delta)}+\frac{1}{w_-(w_++\delta)}\Bigr]\Bigr\},\nonumber\\
\mathcal{D}_{N\Delta}^{\rm C}(\bm{k},
\bm{q})&=&-\frac{1}{4w_+w_-}\Bigl\{\frac{1}{w_++ w_-}
\Bigl[\frac{1}{w_+(w_++\delta)}+\frac{1}{w_-(w_-+\delta)}
\Bigr] \nonumber \\
&~&~~~~~~~~~~~~~~+\frac{1}{\delta+w_++w_-}
\Bigl[\frac{1}{w_++\delta}+\frac{1}{w_-}\Bigr]\Bigl[\frac{1}
{w_-+\delta}+\frac{1}{w_+}\Bigr] \Bigr\}, \nonumber \\
\mathcal{D}_{NN}^{\rm C}(\bm{k}, \bm{q})&=&
-\frac{1}{2w_+w_-}\Bigl\{\frac{1}{w_+w_-}+\frac{1}{w_+^2}+
\frac{1}{w_-^2}\Bigr\}\frac{1}{w_++w_-}\; .
\end{eqnarray}
\end{widetext}
Each of these propagators corresponds also to a different
spin-isospin structure acting in the numerator. However, a
simplification arises from the even parity of the propagators, which
does not affect PNC. Then their angular dependence is only due to
$\cos^2 \theta_{kq}$, which in a good approximation can be taken the
angular average $1/3$ ({\it i.e.} one can replace $\cos\theta_{kq}
\rightarrow \sqrt{1/3}$). Thus PNC actually arises from the vertex
structure of Eqs. \eqref{h1}--\eqref{h3}. In fact, only the terms
with even powers of $k$ survive, when the angular integral is
carried out.

\begin{table}[b]
\caption{\label{tab:params} Parameter values for  meson-nucleon
couplings. The weak couplings $ h_\alpha^{pp},\;
\alpha=\pi,\;\rho,\;\omega$ ($h_\pi^{(1)}$ for pions), are the
"best" estimates of Refs. \cite{ddh} (DDH) and \cite{fcdh} (FCDH).
The strong couplings and cut-offs of the vector mesons are from the
coordinate space OBE of Ref. \cite{mach}.}
\begin{ruledtabular}
\begin{tabular}{cccccc}
& DDH & FCDH \\
&$ h_\alpha^{pp}~(10^{-7})$&$
h_\alpha^{pp}~(10^{-7})$&$g_\alpha^2/4\pi$&$\chi_\alpha$&
$\Lambda_\alpha$~(GeV)\\
\hline
$\pi$ & 4.6 & 2.7 & 13.8 & - & 1.0\\
$\rho$ & $-$15.5 & $-$7.0 & 0.95 & 6.1 & 1.3\\
$\omega$ & $-$3.0 & $-$7.2 & 20 & 0 & 1.5\\
\end{tabular}
\end{ruledtabular}
\end{table}

In momentum space the resulting PNC TPE potential is obtained in the
local form
\begin{equation}\label{pot}
\tilde{V}_{NN2\pi}^{\rm PNC}(\bm{q})=
ih_{\pi}^{(1)}\,(\bm{\sigma}_1\times\bm{\sigma}_2)\cdot
\bm{q}\,\tilde{U}(\bm{q})
\end{equation}
with
\begin{eqnarray}
\label{upot} \tilde{U}(\bm{q})&= \displaystyle
-\frac{4f_\pi^3(\bm{\tau}_1+\bm{\tau}_2)_0}
 {75\sqrt{2}\pi^2m_\pi^3} \;
 \int_0^\infty{dk}\bm{k}^4\Bigl(12\mathcal{D}_{N\Delta}^B-
4\mathcal{D}_{N\Delta}^C+25\mathcal{D}_{NN}^C\Bigr).
\end{eqnarray}
Here we have used the quark model relation $f_\pi^{\ast}$ =
$\sqrt{\frac{72}{25}}f_\pi$ between the strong transition coupling
and the $\pi NN$ coupling \cite{bro}. The isospin operator
contributes a  factor of +2 in the case of elastic proton-proton
scattering.

We calculate Eq. \eqref{upot} executing the integral numerically
with a monopole form factor of the type
\begin{equation}\label{form}
F_{\pm}(\bm{k},\bm{q})=\frac{\Lambda_\pi^2-m_\pi^2}
{\Lambda_\pi^2+\bm{\Pi}_\pm^2}
\end{equation}
included in each vertex.  The parameters given in
Table~\ref{tab:params} are used to calculate $\tilde{U}(\bm{q})$
with $f_\pi = m_\pi g_\pi /2M$, the charged pion mass $m_\pi=139.6$
MeV and the average nucleon mass $M$ = 939 MeV. The relatively small
cut-off mass of the pion is in line (even harder) with the cloudy
bag size arguments \cite{coo,size,tho,hol} and its use here does not
conflict with the Bonn potential parameters used for the vector
mesons.

We fit separately all the three pieces of Eq. \eqref{upot}, which
represent contributions of the different graph types, to see their
relative strengths. Excellent fits (relative error $\le 1$\%) are
achieved with the function of the form
\begin{equation}\label{fit}
\tilde{W}(\bm{q})=A\frac{B^2}{B^2+\bm{q}^2}\Bigl(\frac{C^2}
{C^2+\bm{q}^2}\Bigr)^2.
\end{equation}
with the parameter values given in Table~\ref{tab:fit}.
\begin{table}
\caption{\label{tab:fit} Fit parameters for the Eq. \eqref{part} for
the static partial contributions and the non-static total
contributions in two different kinematics.}
\begin{ruledtabular}
\begin{tabular}{cccc}
&A (fm$^3)$&B (fm$^{-1})$&C (fm$^{-1})$\\
\hline
STATIC\\
$N\Delta^{\rm B}$ & 0.148934 & 3.5987 & 10.8463 \\
$N\Delta^{\rm C}$ & $-$0.0170383 & 3.18247 & 8.28774 \\
$NN^{\rm C}$  & 0.226128 & 2.77668 & 7.59613 \\
\hline
NON-STATIC\\
symmetric & 0.215807 & 2.74463 & 23.6659\\
asymmetric & 0.215388 & 2.13806 & 6.32248
\end{tabular}
\end{ruledtabular}
\end{table}

\begin{figure}
\includegraphics[width=11.0cm]{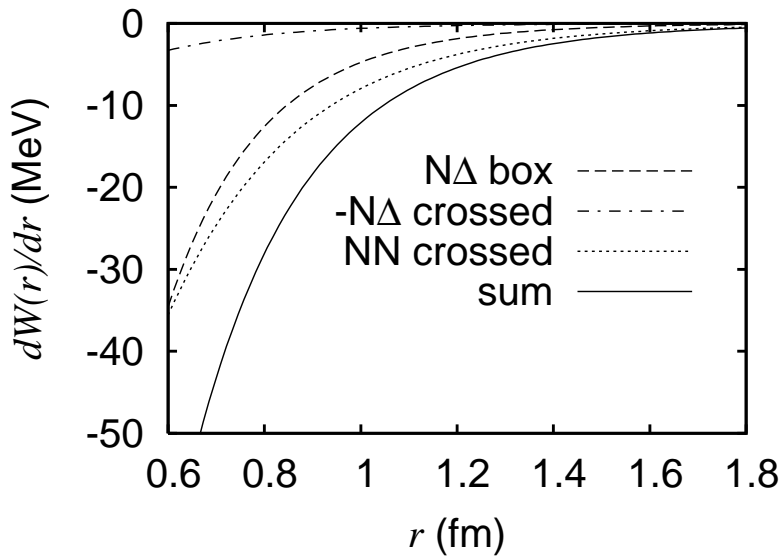}
\caption{\label{figpotr} The total and the partial contributions of the PNC TPE
given by Eq. \eqref{part}.}
\end{figure}

Following  Ref. \cite{csb} we replace Eq. \eqref{upot} with the fit
 in \eqref{pot} and get the configuration space potential as the
 Fourier transform
\begin{equation}\label{potr}
V_{pp2\pi}^{\rm PNC}(\bm{r})=
h_{\pi}^{(1)}(\bm{\sigma}_1\times\bm{\sigma}_2)
\cdot\hat{\bm{r}}\frac{\partial}{\partial r}W(\bm{r})
\end{equation}
with
\begin{eqnarray}\label{part}
\frac{\partial}{\partial r}W(\bm{r})=\frac{AB^2}{4\pi}\Bigl(
\frac{C^2}{C^2-B^2}\Bigl)^2\;
 \Bigl\{e^{-Cr}\Bigl[\frac{C^2-B^2}{2}+
\frac{1}{r}(C+\frac{1}{r})\Bigr]-\frac{e^{-Br}}{r}
(B+\frac{1}{r})\Bigr\}\; .
\end{eqnarray}
The magnitudes of Eq. $\eqref{part}$ for the partial contributions
are illustrated in Fig. \ref{figpotr}, where the $NN$ crossed box
and $\Delta N$ direct box are by far dominant. As expected, the
crossed $NN$ contribution is of the longest range, but the
$\Delta N$ excitation becomes as important or even larger inside
$0.6$ fm. Partly due to the weight factors present in Eq.
\eqref{upot} the crossed $\Delta N$ contribution is nearly negligible.

A comparison with the potential obtained from chiral perturbation
theory \cite{kaiser} may be in order. That contains also the
(reducible) box diagram and a triangle diagram with $s$-wave
rescattering. The former is identically zero in $pp$ scattering
(Barton's theorem), while the latter is chirally suppressed in the
$pp$ case. So our results should be similar. In fact, for distances
larger than about 1.2 fm they are indistinguishable, whereas inside
the radius of 0.8 fm our result is significantly softer due to the
form factors involved. Both have a range corresponding roughly to
vector meson exchanges, so actually probably one may expect similar
results.

\begin{figure}[tb]
\includegraphics[width=11.0cm]{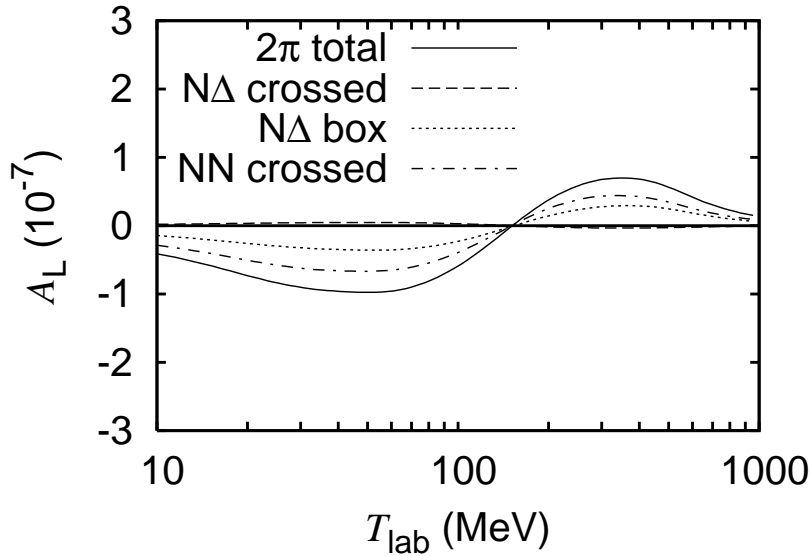}
\caption{\label{Al_partial}  The analyzing power $\bar A_{\rm L}$
arising from the different partial contributions of the PNC 2$\pi$
exchanges using the DDH "best" value for the weak coupling
$h_{\pi}^{(1)}$.}
\end{figure}

We now use the above PNC 2$\pi$ exchange potential to calculate the
asymmetry of the parity-violating spin observable
\begin{equation}\label{al}
\bar{A}_{\rm L} =\frac{\sigma^+-\sigma^-}
{\sigma^++\sigma^-} \; ,
\end{equation}
 where $\sigma^{\pm}$
are the scattering cross sections of the two helicity states of the
longitudinally polarized beam. The Reid soft-core potential
\cite{reid} is used to obtain the strong interaction distortions for
weak interactions, while to minimize theoretical uncertainties
otherwise the empirical phase shifts and strong interaction
amplitudes are used and taken from Ref. \cite{arndt}. It has been
seen in the past that the dependence on the strong potential is
relatively weak and we do not go in this in detail
\cite{dri,carlson}. Therefore, apart from the easily scalable $\pi
NN$ weak coupling and the form factor, the results in Fig.
\ref{Al_partial} may be considered fairly model independent.
Contributions of the  five lowest parity mixed  partial wave
amplitudes ($^1S_0-{}^3P_0$), ($^3P_2-{}^1D_2$), ($^1D_2-{}^3F_2$),
($^1G_4-{}^3F_4$), and ($^1G_4-{}^3H_4$) are included in this
calculation. Structurally the $J=0$ and $J=2$ contributions are
similar to earlier results and the expansion is already converged
for those ($J=4$ is negligible at our energies).

\begin{figure}
\includegraphics[width=11.0cm]{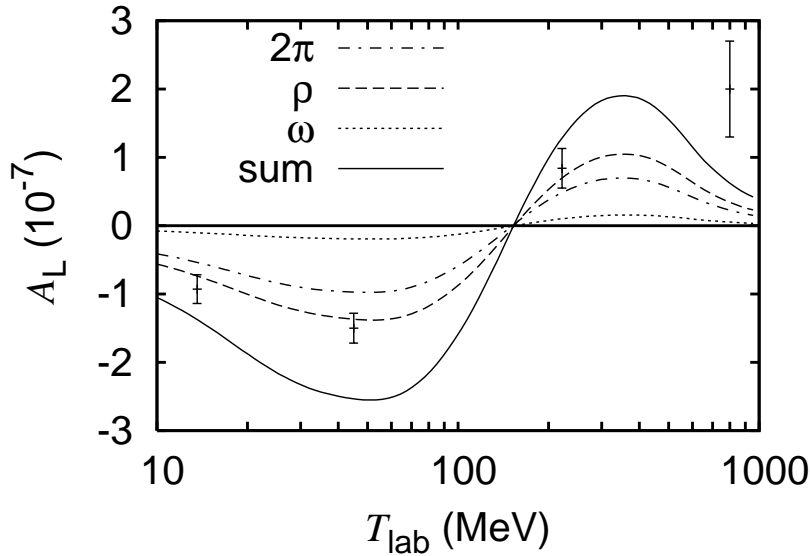}
\caption{\label{al1} Analyzing power $\bar{A}_{\rm
L}$  for the PNC TPE and the local parts of the $\omega$ and $\rho$
exchanges using the DDH "best" values for the weak couplings. The
experimental data points are Bonn at 13.6 MeV \cite{bonn}, PSI at 45
MeV \cite{kis}, TRIUMF at 221.3 MeV \cite{ber}, and Los Alamos at
800 MeV \cite{los}.}
\end{figure}

\begin{figure}[tb]
\includegraphics[width=11.0cm]{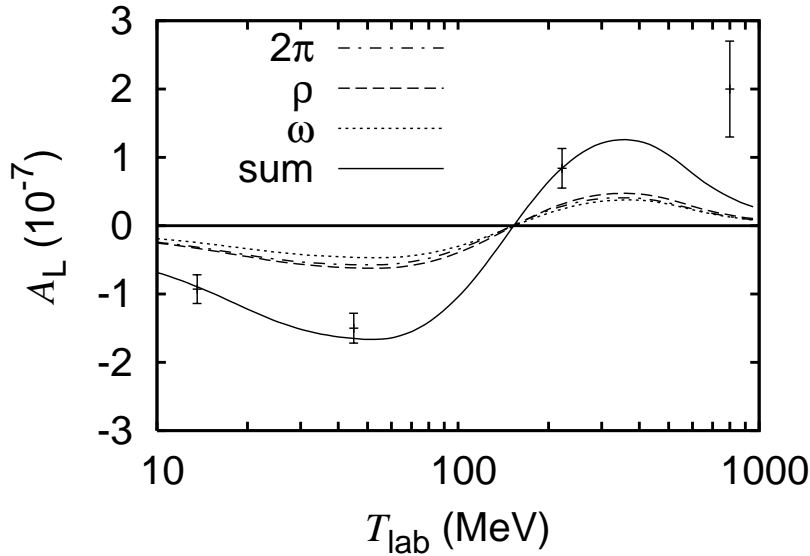}
\caption{\label{al2} The same as Fig. \ref{al1}, except the FCDH
'best' values are employed. The 2$\pi$ and $\rho$ exchanges are now
nearly indistinguishable.}
\end{figure}

As the TPE effect in $\bar A_{\rm L}$ is clearly large, it is of
utmost interest to compare the present result against vector meson
effects considered earlier.  These potentials are given in Ref.
\cite{ddh} in which we also incorporate the monopole form factors of
the type \eqref{form} in each vertex and use two sets of weak
couplings \cite{ddh,fcdh}. Thus the PNC vector meson potentials for $\rho$ and $\omega$ read
\begin{equation}\label{omegapot}
V_{pp\alpha}^{\rm PNC}(\bm{r})=
-\frac{g_\alpha h_{\alpha}^{pp}}{M}\Bigl(
(\bm{\sigma}_1-\bm{\sigma}_2)\cdot\{-i\bm{\nabla},Y_{\alpha}(\bm{r})\}
+  i(1+\chi_\alpha)(\bm{\sigma}_1\times\bm{\sigma}_2)
\cdot[-i\bm{\nabla},Y_{\alpha}(\bm{r})]\Bigr),
\end{equation}
\begin{equation}
\label{modyuk} Y_{\alpha}(\bm{r})=\frac{e^{-m_\alpha r}}{4\pi r}
-\frac{e^{-\Lambda_\alpha r}}{4\pi}\Bigl(\frac{1}{r}
+\frac{\Lambda_\alpha^2-m_\alpha^2}{2\Lambda_\alpha}\Bigr)
\end{equation}
with the relevant parameters given in Table \ref{tab:params}. In the
case of the vector mesons we use only the dominant  local (latter)
term of Eq. \eqref{omegapot} and neglect its nonlocal (former) term,
which only causes a minor contribution on the interaction \cite{dri}
and also does not have a direct correspondence with the present
local interaction. The PNC TPE effect is comparable in size to those
given by vector meson exchanges using both sets (Figs. \ref{al1} and
\ref{al2}). The older set (DDH) would give the sum as an
overestimate, while the new analysis (FCDH) gives a satisfactory
agreement. However, one should remember that the nonlocal PNC would
increase the result somewhat.

\subsection{\label{non-static}Non-static effects}
So far the calculation has totally neglected any kinetic energies of
baryons. This assumption gives simplicity and clarity while still
being probably reasonably realistic due to the large baryon masses.
We now make various attempts to overcome this approximation and to
model non-static effects.

\begin{figure}[tb]
\includegraphics[width=11.0cm]{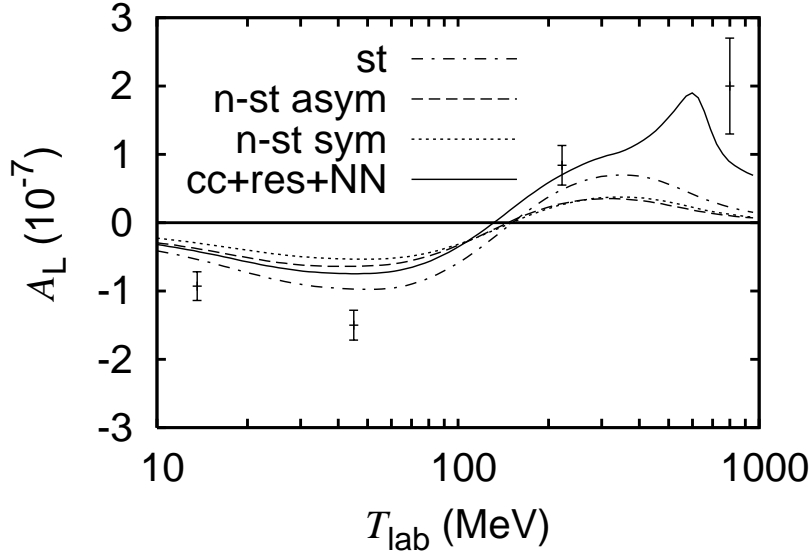}
\caption{\label{last} Model dependence of PNC TPE using the DDH "best"
value for the $h_\pi^{(1)}$. The solid line is
the sum of coupled channels, the residual part, and the $NN$ crossed
contributions. The dash-dotted line is the static, dotted non-static
symmetric, and the dashed non-static asymmetric PNC TPE as discussed
in the text.}
\end{figure}

First we still take the initial kinetic energy (and baryon momenta)
to be zero, while the intermediate and final baryons have nonzero
energies. In this case the final relative momentum is $\bm q$ and
energy is not conserved. In one boson exchange potentials without
internal excitations this is not a problem and presently we are just
making an energy independent potential. The relatively trivial
generalization of Eq. \eqref{props} with $\bm p = \bm q/2$ in Fig.
\ref{kin} yields a qualitatively similar but weaker potential, since
the excitation is larger, and a smaller $\bar A_{\rm L}$
("non-static asymmetric" dashed curve in Fig. \ref{last} and
parametrization in Table \ref{tab:fit}).

Another way of considering non-static effects is to allow kinetic
energy to be present in the initial state with the consequences on
momenta. In this case the assumption of the conservation of energy
would give a simplification making $\bm p$ and $\bm q$ orthogonal
and kinematics could be symmetric. However, in this case numerically
one meets a pole in the integration over $\bm k$ at higher energies,
{\it i.e.} for large incident momenta. Taking for definiteness $\bm
p$ to be zero it is, nevertheless, possible to get a result without
the pole disturbing too much numerics (keeping $q \leq 5\, {\rm
fm}^{-1}$) -- see Fig. \ref{last} (dotted curve) and Table
\ref{tab:fit}.

Probably the best way to include the kinetic energy of the baryons
is to incorporate them dynamically in the equation of motion. This
can be done at the two-baryon level by the coupled-channel
Schr\"odinger equation. Although this does not cover crossed mesons,
following Ref. \cite{csb} it is possible to rearrange the direct and
crossed $\Delta N$ diagrams as
\begin{equation}
\mathcal{N}_{\rm B} \mathcal{D}_{\rm B} +
 \mathcal{N}_{\rm C} \mathcal{D}_{\rm
C} = \mathcal{N}_{\rm B}
 (\mathcal{D}_{\rm B}+\mathcal{D}_{\rm C}) +
  ( \mathcal{N}_{\rm C}- \mathcal{N}_{\rm B}  )\mathcal{D}_{\rm C}\, ,
\end{equation}
where the numerators involving the spin-isospin structure are
denoted by $\mathcal{N}$. The first term is an iteration of static
$\pi$ exchange potentials, since the sum of all propagators turns
out to be just $-(\omega_+^2\,\delta \,\omega_-^2)^{-1}$ and
 $\mathcal{N}_{\rm
B}$ has the corresponding vertex structure. So this term can be
generated by coupled channels iterating an $NN\leftrightarrow \Delta
N$ transition potential, while the second is presumably a smaller
residual correction to be dealt with perturbatively. In terms of the
previous diagrammatic integrals of Eq. \eqref{upot} this
rearrangement means
\begin{eqnarray}\label{CC}
\tilde{U}_{N\Delta}^{\rm CC}(\bm{q})&= &
 \displaystyle  -\frac{32f_\pi^3}{25\sqrt{2}\pi^2m_\pi^3}
\int_0^\infty{dk}\bm{k}^4\Bigl(\mathcal{D}_{N\Delta}^{\rm B}+
\mathcal{D}_{N\Delta}^{\rm C}\Bigr),\\
\label{RES} \tilde{U}_{N\Delta}^{\rm RES}(\bm{q})&=
&\frac{128f_\pi^3}{75\sqrt{2}\pi^2m_\pi^3}
\int_0^\infty{dk}\bm{k}^4\mathcal{D}_{N\Delta}^{\rm C}\; .
\end{eqnarray}
Here the part \eqref{CC} is to be treated in the coupled-channels
approach; the same term would arise from the iteration of the strong
Eq. \eqref{nd} and weak Eq. \eqref{wtrpot} OPE transition
potentials. One advantage of coupled channels is that different
centrifugal barriers related to the orbital angular momenta are
automatically taken into account. Therefore, coupled channels
results are state dependent. Also opening of different channels
causes energy dependence, which cannot be simulated by energy
independent potentials alone.

The first two basic Hamiltonians, Eqs. \eqref{h1} and \eqref{h2},
lead to the strong OPE transition ($NN\leftrightarrow$$N\Delta$)
potential
\begin{eqnarray}\label{nd}
V_{N\Delta\pi}^{\rm PC}(\bm{r})= \frac{f_\pi^\ast
f_\pi}{m_\pi^2}\, \Bigl[(\bm{S}_1\cdot\bm{\nabla})
(\bm{\sigma}_2\cdot\bm{\nabla})\bm{T}_1\cdot\bm{\tau}_2+
(\bm{\sigma}_1\cdot\bm{\nabla})(\bm{S}_2\cdot\bm{\nabla})
\bm{\tau}_1\cdot\bm{T}_2\Bigr]Y_\pi(\bm{r})\, ,
\end{eqnarray}
whereas from  Eqs. \eqref{h2} and \eqref{h3} we get the weak OPE
transition ($NN\leftrightarrow$$N\Delta$) potential
\begin{eqnarray}\label{wtrpot}
V_{\pi N\Delta}^{\rm PNC}(\bm{r})= \displaystyle
\frac{h_\pi^{(1)}f_\pi^{\ast}}{\sqrt{2}m_\pi} \,  \Bigl[
(\bm{\tau}_1\times\bm{T}_2)_0\hat{\bm{r}}\cdot\bm{S}_2+
(\bm{T}_1\times\bm{\tau}_2)_0\hat{\bm{r}}\cdot\bm{S}_1
\Bigr]\frac{\partial}{\partial_r}Y_\pi(\bm{r}).
\end{eqnarray}
Iterating these two should give the effect of the potential
\eqref{CC}, a claim borne out in a comparison of the dashed and
solid (coupled channels + residual + crossed $NN$) curves in Fig.
\ref{last}. At low energies the result is very similar to the
non-static asymmetric potential result but deviates drastically in
energy dependence in particular in the proximity of the $\Delta N$
threshold.

\begin{figure}[tb]
\includegraphics[width=11.0cm]{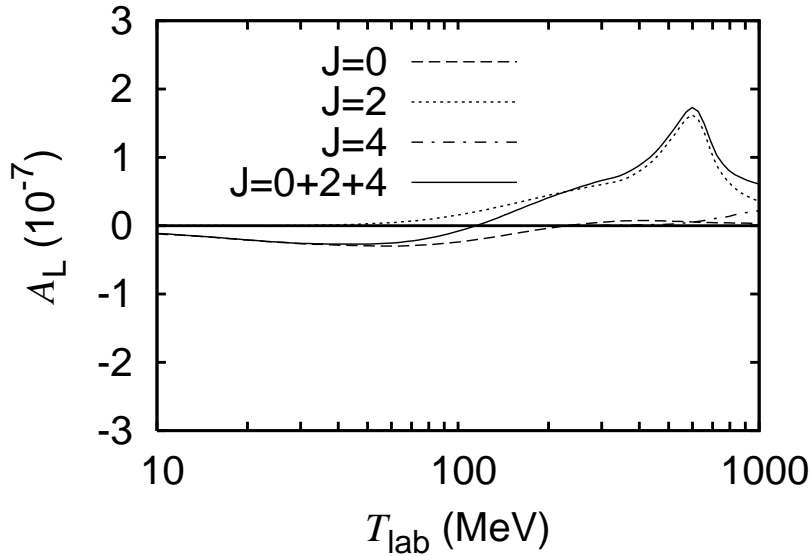}
\caption{\label{iter} Partial wave contributions to $\bar A_{\rm L}$
from PNC TPE using the coupled-channels approach and the DDH "best"
value for the $h_\pi^{(1)}$.}
\end{figure}

It is of interest to study in more detail the dip at 600 MeV in Fig.
\ref{last}. This kind of cusp structures arise in dynamical coupled
channels calculations (as isospin one "dibaryon" resonances in the
case of $\Delta N$ thresholds giving rise to maxima {e.g.} in pion
production), but the present one is somewhat sharper than the wide
maximum seen in Ref. \cite{iq}. However, there it was produced by
PNC $\rho$ exchange interfering with the strong transition involving
both pion and a destructive $\rho$. In the present case, to be
consistent with the TPE potential, we have exclusively the long
ranged pion exchange in both transition potentials without damping
other than the form factor. The structure is due to the favoured
transition $^1D_2(pp) \leftrightarrow\, ^5S_2(\Delta^{++} n)
\leftrightarrow\, ^3P_2(pp)$ through the intermediate state without
a centrifugal barrier. The other intermediate $\Delta N$ states
coupling with tensor-coupled states and with centrifugal barrier are
not particularly favoured \cite{dibar}. These phenomenological
arguments are confirmed in Fig. \ref{iter}, which shows splitting
the above total result into partial wave amplitudes: the structure
is not seen in the otherwise dominant $J=0$ amplitude, which has the
same structure (dictated by strong interaction \cite{simonius}) as
for other potentials.

\section{\label{summ}Conclusion}
We evaluated the PNC TPEP for the elastic $pp$ scattering and
calculated the longitudinal asymmetry $\bar A_{\rm L}$. Compared to
the local contributions of the PNC $\omega$ and $\rho$ exchanges our
results are of the same order indicating that also this mechanism
should be seriously considered in interpreting PNC data. With the
old DDH couplings \cite{ddh} this additional contribution leads to
an overestimate, but the newer ones \cite{fcdh} can give a tolerable
agreement (though we do not include nonlocal PNC here) especially,
if non-static effects are included. However the DDH and FCDH "best"
values for the $h_{\pi}^{(1)}$ might be too large in the light of
the experimental restrictions given by the ${}^{18}$F parity
violating measurements \cite{page,bini}, which bound the upper limit
in the range $|h_{\pi}^{(1)}|$ $\leq 1.5 \times 10^{-7}$. Some
theoretical predictions are also within this limit
\cite{kai,dub,nis}. Further, one should note that the relative sign
of the pion strong and weak coupling may be unknown in the vertex
definitions \eqref{h1}--\eqref{h3} even if the weak magnitude were
given. The knowledge of this coupling is essential, since the pions
are more than five times lighter than the vector mesons and thus the
PNC TPE presumably represents the longest range part of the weak
interaction in $pp$ scattering.

In particular relating to the TRIUMF experiment at 221.3 MeV we get
at that energy the TPE contribution 0.48 to $\bar A_{\rm L}$ in the
static model and in the nonstatic "symmetric" model 0.26 (0.28 in
the "asymmetric" one). These may set rather realistic limits though
the even larger result 0.70 using the coupled channels model should
be noted. However, these are obtained using the older DDH coupling
$h^{(1)}_\pi = 4.6 \cdot 10^{-7}$ and can be easily scaled for the
value from the FCDH analysis $h^{(1)}_\pi = 2.7 \cdot 10^{-7}$ or
any other.

The long ranged OPE transition potential produces also a manifest
cusp peak at 600 MeV in a coupled channels calculation. Whether or
not this will be diminished in the presence of vector mesons will be
discussed in a subsequent work \cite{future}. There are two
competing effects: destructive interference from the $\rho$ exchange
in the strong transition potential and the contribution from the
$\rho$ in the PNC transition as discussed earlier \cite{iq}. In any
case, as suggested in Ref. \cite{iq} an experimental point between
the TRIUMF and Los Alamos energies would be of interest.

\begin{acknowledgments}
We wish to acknowledge the hospitality of Forschungs\-zentrum
J\"{u}lich,  Germany, where this work was partly done and partial
support from the Academy of Finland (grant 121892).
\end{acknowledgments}


\end{document}